\documentstyle[11pt]{article}
\begin{document}

\title{\bf Idealization Second Quantization of Composite Particles}
\author{{D.L. Zhou, S.X. Yu and C.P. Sun}\\
{The Institute of Theoretical Physics, Academia Sinica, Beijing
100080, P.R. China}}
\date{(December 12, 2000)}
\maketitle

\begin{center}
\begin{minipage}{120mm}
\vskip 0.5cm
\begin{center}{\bf Abstract}\end{center}
{A practical method is developed to deal with the second
quantization of the many-body system containing the composite
particles. In our treatment, the modes associated with composite
particles are regarded approximately as independent ones compared
with those of unbound particles. The field operators of the
composite particles thus arise naturally in the second
quantization Hamiltonian. To be emphasized, the second
quantization Hamiltonian has the regular structures which
correspond clearly to different physical processes. }
\end{minipage}
\end{center}

\section*{\bf I. Introduction}

Recently, Feshbach resonances \cite{Feshbach1,Feshbach2} and
photoassociations \cite{photoassociat} in atomic alkli vapours
have renewed one's interest in the many-body systems containing
composite particles. As we know, the second quatization method is
crucial to the description of many-body system, and a many-body
system containing composite particles is universal in nature.
However, the problem on the second quantization of the many-body
system containing composite particles has not been solved
completely in principle. In history, many physicists attempted to
give their projects on it according to their understandings
\cite{classic1,classic2,classic3}. Especially, M.D. Girardeau {\em
et al.} developed the  generalized Fock-Tani transformation
method\cite{focktani1,focktani2,focktani3} which seems to give a
satisfactory solution to this problem. The method is based on
redundant modes and unitary transformation, which is
mathematically strict. However, a subsidiary condition naturally
arises, which prevents using it to solve true many-body problem in
practice. In our opinion, this reflects somehow the impossibility
for the exact second quantization of composite particles.

In addition, the Fock-Tani method is too complex for the computing
of the second quantization Hamiltanian to realistic problem. In
order to obtain the second Hamiltanian easily, the idealization
viewpoint is adopted here that the modes associated with the
unbound particles and the composite particles are independent to
each other, which makes the Fock space constructable. Further, a
projection-operator method is developed to classify the structure
of the Hamiltonian. At last, the second quantization Hamiltonian
is given naturally, which can be verified in the idealization
approximation.

\section*{\bf II. Second Quantization Project}

In this section our second quantization project will be explained
in details with a specific model as an example. This section is
organized as follows. In subsection A, the specific model is
introduced. In subsection B, the Fock space is constructed under a
proper approximation. In the next subsection, the
projection-operator method is introduced to classify the
Hamiltonian in the new picture where
 composite particles are regarded as single entities. Finally, the second
 quantization Hamiltonian is given in subsection D.

\subsection*{\bf A. The Model}

A system composed by $N$ identical bosonic elementary particles is
defined by the Hamiltonian
\begin{equation}
H(1,2,...,N)=\sum_{i=1}^N O(i)+\sum_{i \neq j}T(i,j),
\end{equation}
with $O(i)$ denoting the one-body operator of the $i$-th particle,
$T(i,j)$ denoting the two-body interaction between the $i$-th and
$j$-th particles.

        The composite particles appearing in the system should be defined
according to the Hamiltonian which completely determines the
physical properties of the system. For simplicity, assume that
only two-body composite particles can be formed in the system.
Hence, it is natural to extract the two-body Hamiltonian $H(i,j)$,
which relates to particles $i$ and $j$, from the Hamiltonian
$H(1,2,...,N)$,
\begin{equation}
H(i,j)=O(i)+O(j)+T(i,j).
\end{equation}
However, for physical reasons, it is convenient to select a
Hermitian part $h(i,j)$ of the two-body Hamiltonian $H(i,j)$ (For
how to select $h(i,j)$ in specific situations, see examples in
\cite{explain1}). If the operator $h(i,j)$ admits bound
eigenstates, as we always assuming in the following, the state
vector $|\phi_\alpha(i,j)\rangle$ of the composite particles
forming by particles $i$ and $j$ can be defined to be the bound
eigenstates of $h(i,j)$ , {\em i.e.}
\begin{equation}
h(i,j)|\phi_\alpha(i,j)\rangle=\varepsilon_\alpha
|\phi_\alpha(i,j)\rangle.
\end{equation}

\subsection*{\bf B. The Fock Space}

The essential feature of the second quantization is to express all
the physical quantities and states in Fock space. Our task in this
subsection is to construct the Fock space formally containing
composite particles. However, because the Fock space can not be
constructed strictly, some approximations have to be made in order
to proceed. Here the key assumption is that the Hilbert space of
unbound particles and that of composite particles have the
following ideal properties: I. The Hilbert space of unbound
particles is orthogonal to that of composite particle; II. The
$N$-body Hilbert space can be constructed by the direct product of
$N$ corresponding one-body Hilbert spaces. These properties can be
expressed explicitly in the mathematical languages, i.e. the
orthogonal and complete vector basis of the Hilbert space(without
considering the symmetry requirement) can be chosen as the
following
\begin{equation}
\prod_{m,\alpha} \prod_{t=1}^{n_m} \prod_{s=1}^{n_\alpha}
|\psi_m(m_t)\rangle |\phi_\alpha(\alpha_{s,1},
\alpha_{s,2})\rangle,
\end{equation}
where $\{|\psi_m(i)\rangle, m=1,2,\cdots \}$ are the complete
state vectors of the i-th unbound particle, and correspondingly
$\{|\phi_\alpha(i,j)\rangle, \alpha=1,2,\cdots \}$ are those
vectors of the bound i-th-j-th particle. Note that all the basis
vectors of different configurations together form a complete basis
of the whole Hilbert space.

Thus the basis of the Fock space is defined by
\begin{eqnarray}
|\{n_m(\psi_m)\},\{n_\alpha(\phi_\alpha)\}\rangle&=&
L(\{n_m\}\{n_\alpha\})\sum_P\nonumber\\ &&\prod_{m,\alpha}
\prod_{t=1}^{n_m} \prod_{s=1}^{n_\alpha} |\psi_m(Pm_t)\rangle
|\phi_\alpha(P(\alpha_{s,1}), P(\alpha_{s,2}))\rangle,
\end{eqnarray}
where
\begin{equation}
L(\{n_m\}\{n_\alpha\})=\frac{1}{\sqrt{N! 2^{N_M} \prod_m n_m!
\prod_\alpha n_\alpha!}},
\end{equation}
$ N_A=\sum_m n_m $, $ N_M=\sum_\alpha n_\alpha $, $ N=N_A+2N_M $,
 $\sum_P$ is a sum over all
permutation of $N$ index of particles.

Obviously the above basis vectors have ideal properties as follows
\begin{equation}
\langle \{n^\prime_m\} \{n^\prime_\alpha\}|\{n_m\} \{n_\alpha\}
\rangle=\prod_{m,\alpha} \delta_{n^\prime_m,n_m}
\delta_{n^\prime_\alpha,n_\alpha}.
\end{equation}

Based on the Fock space defined as above, the creation operators
and annihilation operators of the independent mode can be defined
as usual
\begin{eqnarray}
a_m=\sum_n \sqrt{n} |(n-1)_m\rangle \langle n_m|, \\ a_\alpha=
\sum_n \sqrt{n} |(n-1)_\alpha \rangle \langle n_\alpha|.
\end{eqnarray}
Obviously, they obey ideal Bose commutation relations
\begin{equation}
[a_m,a^\dag_l]=\delta_{ml},
[a_\alpha,a^\dag_\beta]=\delta_{\alpha\beta}.
\end{equation}

\subsection*{\bf C. Projected Hamiltonian}

In this subsection, a project operator method is developed to
obtain Hamiltonian in terms of operators including that of
composite particles. The advantage of the project operator method
is that it is convenient to classify the interactions according to
unbound particles and composite particles.

The project operator $S(i)$ for the unbound particle $i$ is
\begin{equation}
S(i)=\sum_m|\psi_m(i)\rangle\langle\psi_m(i)|.
\end{equation}

Correspondingly, the project operator $C(i,j)$ for the composite
particle composed by particle i and j is
\begin{equation}
C(i,j)=\sum_\alpha
|\phi_\alpha(i,j)\rangle\langle\phi_\alpha(i,j)|.
\end{equation}
Note that the project operator $C(i,j)=C(j,i)$.

The unit project operators of $N$ particles is
\begin{eqnarray}
I(1,2,\cdots,N)&&=\sum_{N_A+2N_M=N}\frac 1 {N_A!2^{N_M}N_M!}
\sum_P S(PA_1)\cdots S(PA_{N_A})\nonumber
\\&&C(P(M_{1,1}),P(M_{1,2}))\cdots C(P(M_{N_M,1}),P(M_{N_M,2})).
\end{eqnarray}
In fact, the above equation can follows directly from the key
assumptions as discussed above.

Applying the project operators, the Hamiltonian can be rewrite as
\begin{eqnarray}
H(1,2,\cdots,N)&=&I(1,2,\cdots,N)H(1,2,\cdots,N)I(1,2,\cdots,N)\nonumber\\
&=&H_{SS}(1,2,\cdots,N)+H_{SSSS}(1,2,\cdots,N)+H_{CC}(1,2,\cdots,N)
\nonumber\\
&&\mbox{}+H_{CSS}(1,2,\cdots,N)+H_{SSC}(1,2,\cdots,N)\nonumber\\
&&\mbox{}+H_{SCSC}(1,2,\cdots,N) +H_{CCCC}(1,2,\cdots,N)+\cdots,
\end{eqnarray}
where
\begin{eqnarray}
H_{SS}(1,2,\cdots,N)&=& \sum_{i=1}^N S(i)O(i)S(i),\\
H_{SSSS}(1,2,\cdots,N)&=&\sum_{i<j}S(i)S(j)T(i,j)S(i)S(j),\\
H_{CC}(1,2,\cdots,N)&=&\sum_{i<j}C(i,j)(O(i)+O(j)+T(i,j))C(i,j),\\
H_{CSS}(1,2,\cdots,N)&=&\sum_{i<j}C(i,j)(O(i)+O(j)+T(i,j))S(i)S(j),\\
H_{SSC}(1,2,\cdots,N)&=&\sum_{i<j}S(i)S(j)(O(i)+O(j)+T(i,j))C(i,j),\\
H_{SCSC}(1,2,\cdots,N)&=&\sum_{i\neq j<k}
S(i)C(j,k)(T(i,j)+T(i,k))S(i)C(j,k)\nonumber\\
 &&\mbox{}+\sum_{i\neq j<k}(S(j)C(i,k)+S(k)C(j,i))(O(i)+O(j)+O(k)\nonumber\\
 &&\mbox{}+T(i,j)+T(j,k)+T(k,i))S(i)C(j,k),\\
H_{CCCC}(1,2,\cdots,N)&=&
\sum_{i<j{\neq}k<l}C(i,j)C(k,l)(T(i,k)+T(i,l)+T(j,k)+T(j,l))\nonumber\\
&&C(i,j)C(k,l)+\sum_{i<j\neq
k<l}(C(i,k)C(j,l)+C(i,k)C(j,l))\nonumber\\
&&(O(i)+O(j)+O(k)+O(l)+T(i,j)+T(k,l)+T(i,k)\nonumber\\
&&\mbox{}+T(i,l)+T(j,k)+T(j,l))C(i,j)C(k,l).
\end{eqnarray}

Note that in the above equations, only interactions including up
to two-body have been considered if  a composite particle is
regarded as one entity. Compared with systems without composite
particles, the main feature is that it includes the rearrange
terms $H_{CSS}$ $H_{SSC}$. In addition, the Hamiltonian including
infinite terms even if the pre-quantization Hamiltonian only
contain two-body interactions.

\subsection*{\bf D. The Second Quantization Hamiltonian}

Based on the results obtained in the above subsection, the second
quantization Hamiltonian including composite particles will be
given directly as follows.
\begin{eqnarray}
H_{SS}&=&\sum_{m,n}a^\dag_n \langle\psi_n(1)|O(1)|\psi_m(1)\rangle
a_m,\\ H_{SSSS}&=& \frac 1 2 \sum_{m,n,p,q} a^\dag_m
a^\dag_n\langle\psi_m(1) \psi_n(2)|T(1,2)|\psi_p(2)
\psi_q(1)\rangle a_p a_q,\\
H_{CC}&=&\sum_{\alpha,\beta}a^\dag_\alpha
\langle\phi_\alpha(1,2)|O(1)+O(2)+T(1,2)|\phi_\beta(1,2)\rangle
a_\beta,\\ H_{CSS}&=&\frac 1
{\sqrt{2}}\sum_{\alpha,m,n}a^\dag_\alpha
\langle\phi_\alpha(1,2)|O(1)+O(2)\nonumber\\
&&\mbox{}+T(1,2)|\psi_m(2) \psi_n(1)\rangle a_m a_n,\\
H_{SSC}&=&\frac 1 {\sqrt{2}}\sum_{m,n,\alpha}a^\dag_m a^\dag_n
\langle\psi_m(1) \psi_n(2)|O(1)+O(2)\nonumber\\
&&\mbox{}+T(1,2)|\phi_\alpha(1,2)\rangle a_\alpha,\\
H_{SCSC}&=&\sum_{m,n,\alpha,\beta} a^\dag_m a^\dag_\alpha
[\langle\psi_m(1)
\phi_\alpha(2,3)|T(1,2)+T(1,3)|\phi_\beta(2,3)\psi_n(1)\rangle\nonumber\\
&& \mbox{}+\langle\psi_m(1) \phi_\alpha(2,3)|O(1)+O(2)+O(3
)+T(1,2)+T(1,3)\nonumber\\ &&
\mbox{}+T(2,3)|\phi_\beta(1,3)\psi_n(2)\rangle+\langle\psi_ m(1)
\phi_\alpha(2,3)|O(1)+O(2) \nonumber\\ &&
\mbox{}+O(3)+T(1,2)+T(1,3)+T(2,3)|\phi_\beta(1,2)\psi_n(3)\rangle]a_\beta
a_n,\\ H_{CCCC}&=& \frac 1 2
\sum_{\alpha,\beta,\theta,\tau}a^\dag_\alpha a^\dag_\beta
[\langle\phi_\alpha(1,2)\phi_\beta(3,4)|T(1,3)+T(1,4)
+T(2,3)\nonumber\\ &&
\mbox{}+T(2,4)|\phi_\theta(3,4)\phi_\tau(1,2)\rangle+\langle\phi_\alpha(1,2)
\phi_\beta(3,4)|O(1)+O(2)\nonumber\\ &&\mbox{}+O(3)+O(4)
+T(1,2)+T(1,3)+T(1,4)+T(2,3)+T(2,4)\nonumber\\
&&\mbox{}+T(3,4)|\phi_\theta(2,4)\phi_\tau(1,3)\rangle
+\langle\phi_\alpha(1,2)\phi_\beta(3,4)|O(1)+O(2)\nonumber\\
&&\mbox{}+O(3)+O(4) +T(1,2)+T(1,3)+T(1,4)+T(2,3)\nonumber\\
&&\mbox{}+T(2,4)+T(3,4)|\phi_\theta(2,3)\phi_\tau(1,4)\rangle]
a_\tau a_\theta.
\end{eqnarray}

Note that the terms of the above second quantization Hamiltonian
containing composite particles have regular structures. In the
picture in which unbound particles and composite particles are
both regarded as single entities, the terms $H_{SS}$ and $H_{CC}$
are one-body operators; the terms $H_{SSSS}$, $H_{CCCC}$ and
$H_{SCSC}$ are two-body operators, which describe the interactions
between two unbound particles, between two composite particles,
and between one unbound particle and one composite particle
respectively; the terms $H_{CSS}$ and $H_{SSC}$ represent the
rearrangement between two unbound particles and one composite
particle, which don't conserve the total particle number in the
sense that one composite particle as one particle. we also notice
that the coefficient of every term in the Hamiltonian is regular
in the sense that the project operator method is valid.

Of course, the mathematical derivation is needed, which is in fact
quite compact, and will be demonstrated in Appendix.

\section*{\bf III. Discussions and Conclusions}

In Sec.II, our project of second quantization of composite
particles has been applied to a system of $N$ identical bosonic
particles. It can be generalized to cases including $N$ identical
bosonic particles or $N_1$ identical bosonic particles and $N_2$
identical fermionic particals. Also it can be generalized to
composite particles composed by three particles,or more particles.

It should be emphasized here that our project is based on the
approximation that the modes associated with unbound and composite
particles are independent modes. Strictly speaking this is not the
case, therefore theoretical difficulties arise in constructing the
Fock space and second quantization is impossible in principle.
However, if the approximation is acknowledged in idealization
sense, our project is a natural one to construct the second
quantization Hamiltonian of composite particles. Our final results
show that the second quantization Hamiltonian have regular
structures indeed, which is not explicitly given in the classic
papers \cite{classic1,classic2,classic3}.

\vskip 0.5cm {\bf Acknowlegement}:  One of the authors(D.L. Zhou)
thanks for helpful discussions with Dr. X.X. Yi and Dr. Y.X. Liu.
This work is supported by NSF of China.

\vskip 0.5cm
\appendix
\section*{\centerline {Appendix}}
\subsection*{Derivation of the Second Quantization Hamiltonian}

In this appendix, we will give a typical example to demonstrate
how to derive the second quantization Hamiltonian.

\begin{eqnarray}
\ \ \ \ \ \ \ &H_{CSS}&|\{n_m\} \{n_\alpha\}\rangle \equiv
H_{CSS}(1,2,\cdots,N)|\{n_m\} \{n_\alpha\}\rangle \nonumber\\
&=&\sum_{i<j}C(i,j)(O(i)+O(j)+T(i,j))S(i)S(j)\nonumber\\&&
L(\{n_m\}\{n_\alpha\})\sum_P \prod_{m,\alpha} \prod_{t=1}^{n_m}
\prod_{s=1}^{n_\alpha} |\psi_m(Pm_t)\rangle
|\phi_\alpha(P(\alpha_{s,1}), P(\alpha_{s,2}))\rangle\nonumber\\
&=&\sum_{\beta,p,q}\frac 1 {2!}\sum_{i\neq j}
|\phi_\beta(i,j)\rangle\langle\phi_\beta(i,j)|O(i)+O(j)+T(i,j)|\psi_p(i)
\psi_q(j)\rangle\langle\psi_p(i)\psi_q(j)|\nonumber\\ &&
L(\{n_m\}\{n_\alpha\})\sum_P \prod_{m,\alpha} \prod_{t=1}^{n_m}
\prod_{s=1}^{n_\alpha} |\psi_m(Pm_t)\rangle
|\phi_\alpha(P(\alpha_{s,1}), P(\alpha_{s,2}))\rangle\nonumber\\
&=&\frac 1
{2!}\sum_{\beta,p=q}\langle\phi_\beta(1,2)|O(1)+O(2)+T(1,2)|\psi_p(1)
\psi_q(2)\rangle\nonumber\\&&
n_p(n_p-1)L(\{n_m\}\{n_\alpha\})\sum_P \prod_{m,\alpha}
\prod_{t=1}^{n_m^\prime} \prod_{s=1}^{n_\alpha^\prime}
|\psi_m(Pm_t)\rangle |\phi_\alpha(P(\alpha_{s,1}),
P(\alpha_{s,2}))\rangle\nonumber\\ &&\mbox{}+\frac 1
{2!}\sum_{\beta,p\neq q}
\langle\phi_\beta(1,2)|O(1)+O(2)+T(1,2)|\psi_p(1)
\psi_q(2)\rangle\nonumber\\&& n_p n_q L(\{n_m\}\{n_\alpha\})\sum_P
\prod_{m,\alpha} \prod_{t=1}^{n_m^{\prime\prime}}
\prod_{s=1}^{n_\alpha^{\prime\prime}} |\psi_m(Pm_t)\rangle
|\phi_\alpha(P(\alpha_{s,1}), P(\alpha_{s,2}))\rangle\nonumber\\
&=&\frac 1
{\sqrt{2!}}\sum_{\beta,p=q}\sum_P\langle\phi_\beta(1,2)|O(1)+O(2)+T(1,2)|\psi_p(1)
\psi_q(2)\rangle\sqrt{n_\beta+1}\nonumber\\&&\sqrt{n_p(n_p-1)}
L(\{n_m^\prime\}\{n_\alpha^\prime\})\sum_P \prod_{m,\alpha}
\prod_{t=1}^{n_m^\prime} \prod_{s=1}^{n_\alpha^\prime}
|\psi_m(Pm_t)\rangle |\phi_\alpha(P(\alpha_{s,1}),
P(\alpha_{s,2}))\rangle\nonumber\\ &&\mbox{}+\frac 1
{\sqrt{2!}}\sum_{\beta,p\neq
q}\sum_P\langle\phi_\beta(1,2)|O(1)+O(2)+T(1,2)|\psi_p(1)
\psi_q(2)\rangle\sqrt{n_\beta+1}\nonumber\\&&\sqrt{n_p
n_q}L(\{n_m^{\prime\prime}\}\{n_\alpha^{\prime\prime}\})\sum_P
\prod_{m,\alpha} \prod_{t=1}^{n_m^{\prime\prime}}
\prod_{s=1}^{n_\alpha^{\prime\prime}} |\psi_m(Pm_t)\rangle
|\phi_\alpha(P(\alpha_{s,1}), P(\alpha_{s,2}))\rangle\nonumber\\
&=&\frac 1 {\sqrt{2!}}\sum_{\beta,p,
q}a^\dag_\beta\langle\phi_\beta(1,2)|O(1)+O(2)+T(1,2)|\psi_p(1)
\psi_q(2)\rangle a_p a_q|\{n_m\} \{n_\alpha\}\rangle.
\end{eqnarray}
where
\begin{eqnarray}
n_m^\prime &=& \left \{
                \begin{array}{ll}
                n_m-2 & {\rm if\ } m=p=q\\
                n_m   & {\rm otherwise}
                \end{array}
                \right.,\\
n_\alpha^\prime &=& \left \{
                \begin{array}{ll}
                n_\alpha+1 & {\rm if\ } \alpha=\beta\\
                n_\alpha   & {\rm otherwise}
                \end{array}
                \right.,\\
n_m^{\prime\prime} &=& \left \{
                \begin{array}{ll}
                n_m-1 & {\rm if\ } m=p\ or\ q\\
                n_m   & {\rm otherwise}
                \end{array}
                \right.,\\
n_\alpha^{\prime\prime} &=& \left \{
                \begin{array}{ll}
                n_\alpha+1 & {\rm if\ } \alpha=\beta\\
                n_\alpha   & {\rm otherwise}
                \end{array}
                \right..
\end{eqnarray}
We thus have
\begin{equation}
H_{CSS}=\frac 1 {\sqrt{2!}}\sum_{\beta,p,
q}a^\dag_\beta\langle\phi_\beta(1,2)|O(1)+O(2)+T(1,2)|\psi_p(1)
\psi_q(2)\rangle a_p a_q.
\end{equation}

In fact, the similar procedure as above can be used to obtain the
other terms.


\begin{thebibliography}{99}
\bibitem {explain1} For examples, external potential,
if it exists, often should be excluded in the definition of
composite particles; In addition, for resonance states, the weak
interactions between the discrete state and the continuum states
may be ignored in this process.
\end{thebibliography}
\end{document}